The impact of small changes in thoroughfare connectivity on the potential for student walking

Jeremy D. Auerbach[1], Eugene C. Fitzhugh[2], and Ellen Zavisca[3]

[1] Department of Environmental and Radiological Health Sciences
Colorado State University, Fort Collins, CO 80523, USA

[2] Department of Kinesiology, Recreation, and Sport Studies
University of Tennessee, Knoxville, TN 37996, USA

[3] Knoxville Regional Transportation Planning Organization
City of Knoxville, TN, 37902, USA

Corresponding author information:
Jeremy D. Auerbach
Department of Environmental and Radiological Health Sciences
Colorado State University, Fort Collins, CO 80523, USA
Phone: 970-491-1063
Fax: 970-491-2940
E-mail: jeremy.auerbach@colostate.edu

**Abstract**

**Introduction.** Student active commuting to school is an important component to student achievement and student health, yet this form of physical activity has significantly declined in the U.S. Distance between the school and student residence is often reported as a barrier for student walking, thereby increasing street and trail connectivity between and within residential developments and schools could foster student walking. The purpose of this study is to evaluate the potential benefits of increased thoroughfare connectivity on student walking within school walking zones.

**Methods.** This study conducts a cost-benefit analysis of increased thoroughfare connectivity around elementary and middle schools in a U.S. school system that serves sixty thousand students. Benefits, which include the increased time of physical activity from student walking and the potential cost-savings to a school system if they had fewer students to bus to school, are compared to the financial costs of the new connections. Advanced network optimization techniques were applied to several suburban and rural schools from a representative the school system to locate the optimal new thoroughfare connections that maximize student walking to a school and minimize the length of the new thoroughfare.

**Results.** Results from this case study showed that short and inexpensive new thoroughfares could increase the potential number of student active commuters and provide a significant increase of physical activity for those potential student walkers.

**Conclusions.** This work can foster the integration of student walking and student health in residential planning decisions around schools.

## 1. Introduction

There has been a rapid decline in physical activity reported for U.S. children (Nader et al., 2008). This lack of physical activity is one of the primary contributors of childhood obesity and physical activity is a valuable component of childhood development and academic achievement regardless of socioeconomics, demographics, geography, and school characteristics (Robert Wood Johnson Foundation, 2009, Centers for Disease Control and Prevention, 2010). Increased physical activity has also been correlated with higher attendance rates and fewer disciplinary incidents (Welk, 2009).

One way to combat this trend of declining physical activity in children is to promote student active commuting to school, through walking or biking. Children who actively commute to school tend to be more active outside of this commuting to school and active commuting to school has been shown to be inversely associated with body mass index (Sirard and Slater, 2008, Lubans et al., 2011, Mendoza et al., 2011, Turrell et al., 2018). Not surprisingly, with the decline of physical activity in children in the U.S., there has also been a decline of active commuting to school (McDonald, 2007, Pedestrian and Bicycle Information Center, 2010, The National Center for Safe Routes to School, 2016). Parents have reported several behavioral and environmental barriers that have contributed to this decline: the perception of possible violence or crime along the route; the speed and volume of traffic along the route; poor weather or climate in the area; and the distance between home and school (Nelson et al., 2008, Pedestrian and Bicycle Information Center, 2010, McDonald and Aalborg, 2009, Centers for Disease Control and Prevention, 2005, Mendoza et al., 2014).

In the U.S., centrally located school have been advocated to promote student walking since at least the 1920's (Perry, 1929). More centrally located schools have more students walking or biking, even after controlling for other neighborhood characteristics (Kim and Lee, 2016), but recent trends in the school site design and location in the U.S. make active transportation difficult (Chriqui et al., 2012). Since the 1970s, school systems have increasingly constructed larger schools on larger tracts of land, often in rural areas further from population centers with the number of schools dropping from 262,000 in 1930 to 95,000 in 2004, while the student population has increased from 28M to 54M (Office of Children's Health Protection, 2011).

With these barriers and trends in the siting of U.S. schools not changing another avenue is to increase the walkability of the residential neighborhoods proximate to schools. This connectivity of the built environment has been found to be positively correlated with student active commuting as it contributes to a shorter distance between school and home (Frank et al., 2006, Babey et al., 2009, Bungum et al., 2009, Larsen et al., 2009, Giles-Corti et al., 2011, Coughenour et al., 2017, Marshall et al., 2014). Yet, the neighborhood design features that are currently popular with residents are at odds with thoroughfare connectivity. Residential neighborhood developers are incentivized to minimize thoroughfare connectivity in order to reduce the number of community entrances and increase the number of buildable lots, especially lots on cul-de-sacs. These built environment features are desirable for residents as they provide low-traffic streets and the perception of security associated with them. As expected, reduced neighborhood walkability and increased distance between home and school has been shown to be positively correlated with childhood obesity (Spence et al., 2008, Grafova, 2008, Oreskovic et al., 2009, Tewahade et al., 2019).

This decline in active commuting to schools also leads to increased traffic from private driving and school busing, which accounts for 10-14% of morning rush-hour traffic (McDonald et al., 2011), and increased pollution. The additional traffic intensifies parental safety concerns related to traffic and student active commuting. Still, the number of children killed and injured while walking or biking is dwarfed by the increasing rate of vehicle crashes, which are the leading cause of death among school age children in the U.S. (National Center for Statistics and Analysis, 2016). Furthermore, this decrease in active commuting also imposes a considerable economic burden on school systems with increased costs related to purchasing and maintaining buses, hiring drivers (an occupation with a high turnover rate), and fuel and insurance expenditures (DeNisco, 2015). A study of school administrators found that addressing the perceived safety concerns and increasing the number of sidewalks can increase active travel to schools (Price et al., 2011), and one the key elements of the National Center for Safe Routes to School is transportation planning approaches to ensure safe active commuting opportunities (The National Center for Safe Routes to School, 2016).

These school siting trends and neighborhood design features contribute to urban sprawl and to greater distances between student homes and schools, and when coupled with a lack of sidewalks and bike paths greatly reduce student opportunities for active commuting (Kouri, 1999). The purpose of this study is to estimate the potential benefits of designing residential developments with more thoroughfare connectivity. In order to do that, it models how existing walking zones could be expanded via greater thoroughfare connectivity around schools. Several potential benefits may result from this research: (1) the study will examine the potential cost-savings to school systems if they had to provide busing for a smaller proportion of students, (2) the expected health impact to students (e.g., increased physical activity) by increasing the potential that they will walk or bike to school, and (3) an approachable methodology to automate the discovery of potential thoroughfares for planners, school officials, and researchers.

## 2. Methods

### 2.1 Case Study

We have worked closely with the Knox County School (KCS) public school system to evaluate the impact of increased thoroughfare connectivity on student walking. KCS serves 60,000 students in and surrounding Knoxville, Tennessee, with 89 schools and 337 buses. School acreage and attendance growth for schools in the KCS have undergone changes similar to those experienced by school systems across the US (Transportation Consultants, 2014). The average acreage of elementary schools jumped from 8.5 acres for schools built prior to 1977 to 24.5 acres for schools built since then (see Fig. 1). Due to school site selection in Knox County, from the 2013-14 school year to the 2017-18 school year, the number of students who lived outside the walking distance increased by approximately 6,000. As with most US school systems, KCS has an established policy to determine if a student is eligible for transportation by a bus, based on each student's residence location in relation to their school (Knox County Schools Transportation Department, 2009). A student is inside the school walk zone (SWZ), and ineligible for busing, if the distance from the student's residence to the drop-off location of the student's zoned school is less than 1 mile (for elementary students) or less than 1.5 miles (for secondary students), based upon the existing street network.

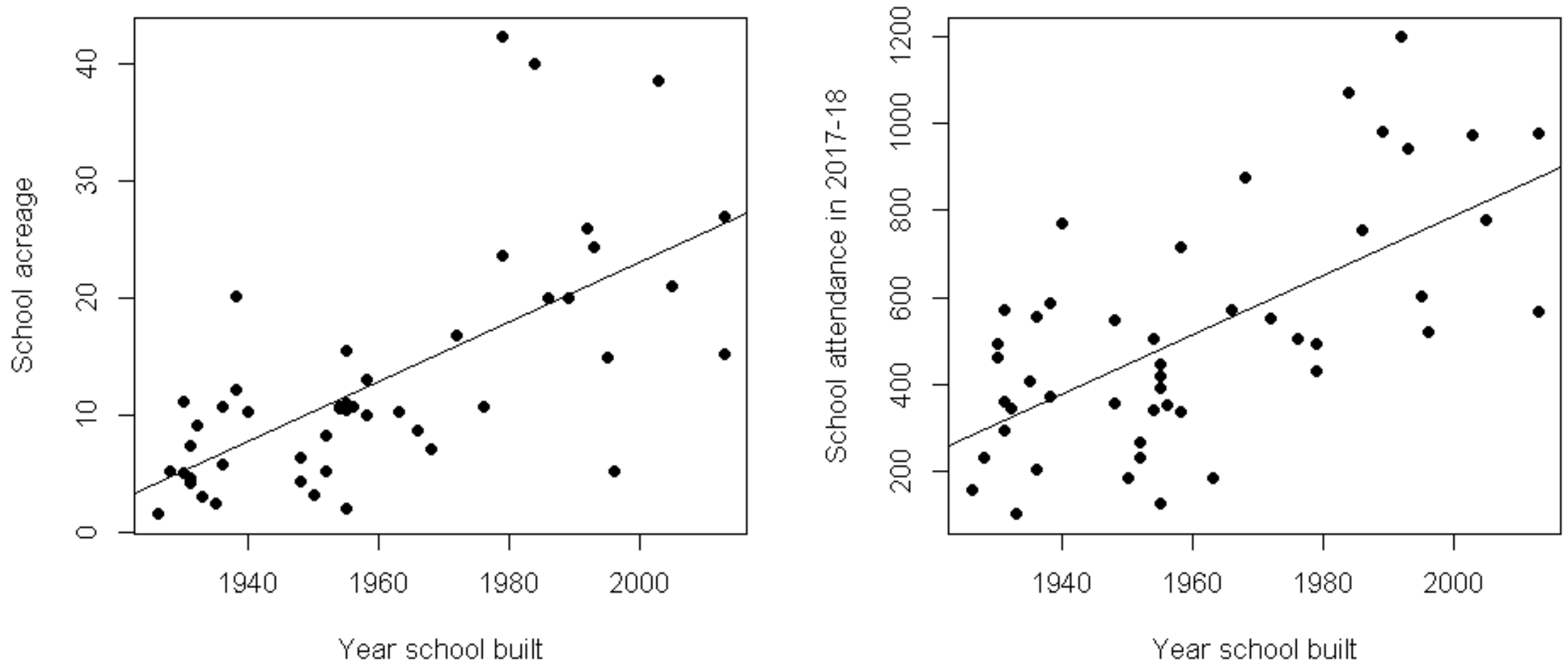

Fig. 1. School acreage and attendance growth for schools in the Knox County School System.

A report by the Knox County Department of Engineering and Public Works identified that 37.6% of the total number of students in the school district (22,322 students) lived within the SWZs (Transportation Consultants, 2014). They calculated the shortest path distance for these students, based on the thoroughfare topology, and found more than 30% of those walking distances were longer than the SWZ distance. Therefore, KCS is responsible for busing large numbers of students (up to 7,434 students) to school who could potentially be within active transportation distance, which places a significant financial burden on the school district. A significant contributor to this to distance from residence to school is poorly connected street networks. An analysis conducted by the authors of this study found many Knox County schools were near neighborhoods with street networks that had few intersections or frequent dead-ends (cul-de-sacs), i.e. neighborhoods with low thoroughfare connectivity. As such, the network SWZs fail to capture many students that reside close to the school “as the crow flies," that is, the straight-line or “Euclidean distance" (see Fig. 2.A).

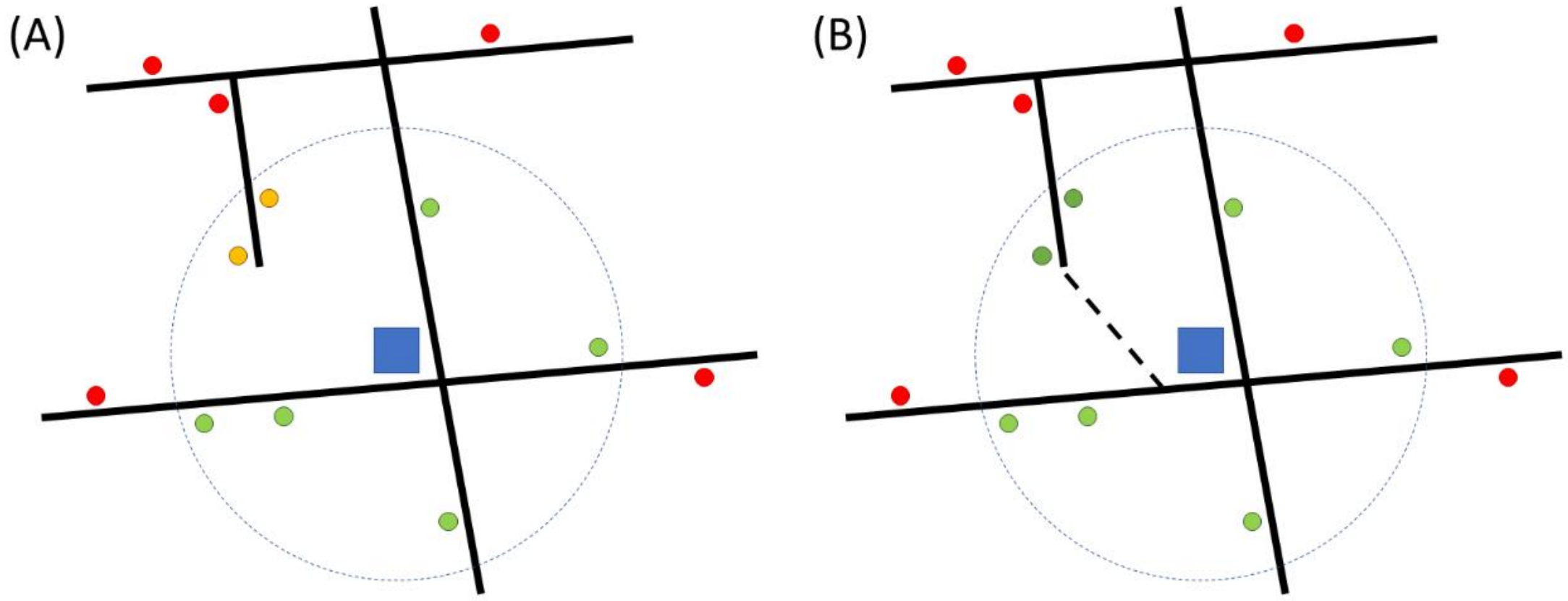

Fig. 2. An example of the benefit of a new thoroughfare connection. In figure A orange nodes represent the residential parcels within the Euclidean SWZ distance (blue dashed line) of the school (blue square) but not within the network SWZ distance (black lines denote streets). In figure B, the residential parcels within the network SWZ distance are represented by green nodes and the residences outside of the Euclidean SWZ distance are the red nodes.

## 2.2 Data

Since student residences for a given school can change over the course of the year and from year to year, as students enroll, graduate, or move, we used residential parcels as the proxy for student residence locations. Data about residential parcel locations and types, such as single-family residence or multi-family residence, were provided by GIS administrator for Knoxville, Knox County, and the Knoxville Utilities Board. Knoxville-Knox County Planning has provided the average number of students for residential parcel types in the study area, which was used to estimate the number of students that would be affected by changes in thoroughfare connectivity, the number of students for each school in Knox County, the number of students within each school's SWZ, and the number of students that are within the SWZ and network distance to their respective school. The data was collected in 2017 and engineered according to the methods provided in the Supplemental Materials.

## 2.3 School Selection

Ten schools that would benefit the most from additional thoroughfare connectivity were selected for analysis with the SWZ distance disparity metric. SWZ distance disparity was developed for this study and is the difference between the number of students in the Euclidean SWZ distance and the number of students in the network SWZ. A large proportion of students within both the Euclidean SWZ distance and the network SWZ is associated with an environment that has high thoroughfare connectivity, typically urban. A large proportion of students within the Euclidean SWZ distance and a low proportion in the network SWZ is indicative of a built-up yet low-connectivity environment, typically suburban. Low proportions in both the Euclidean SWZ distance and the actual SWZ is associated with a low-density, typically rural, environment. The ten schools with the largest SWZ distance disparity were selected for the analysis in this study (see Fig. 3.). These ten schools were from suburban and rural environments, as urban schools already had large proportions of students within the network SWZs and therefore low SWZ distance disparity.

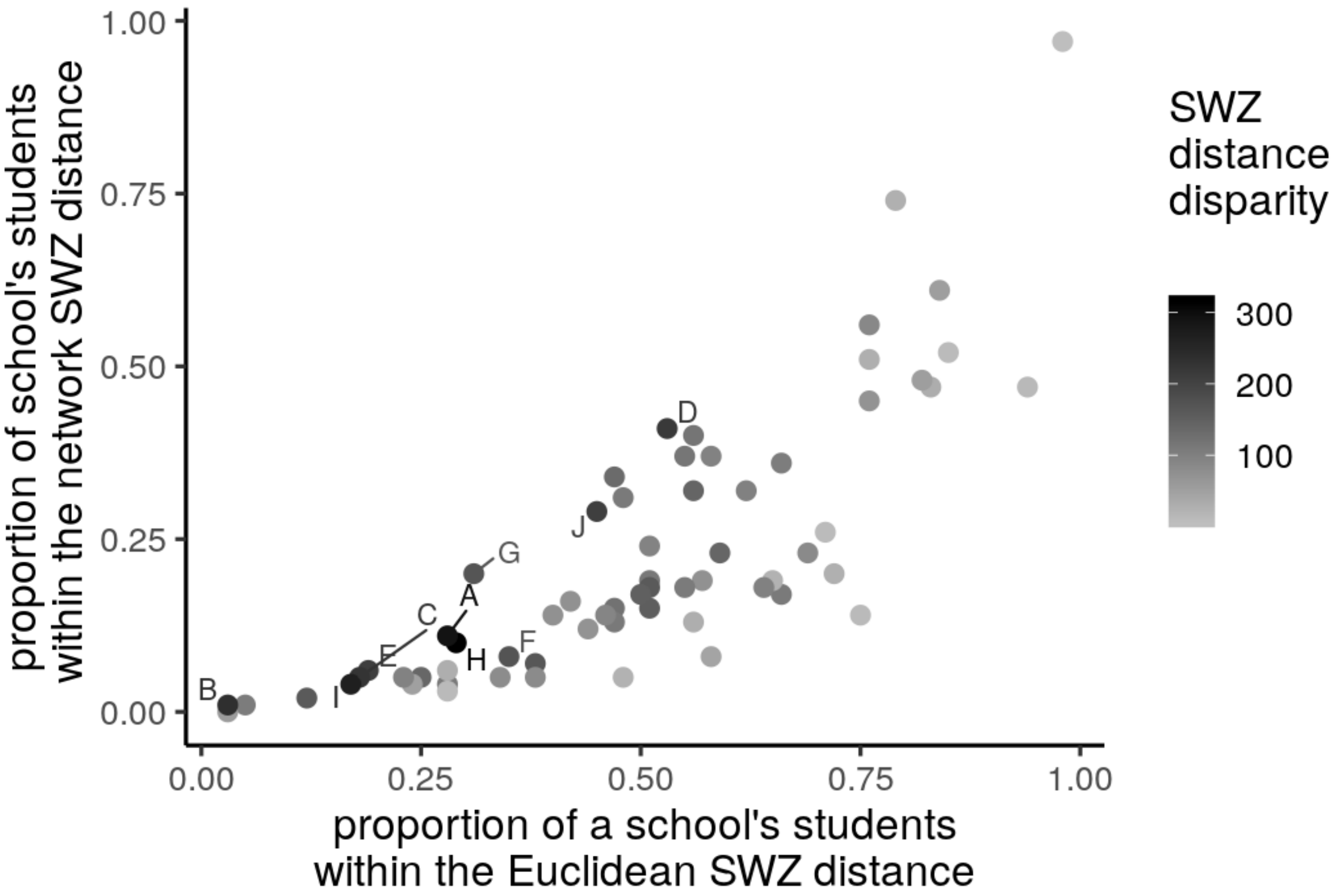

Fig. 3. SWZ distance disparity for Knox County schools. The ten schools with highest SWZ distance disparity, i.e. the number of students who would be in the SWZ if there were additional street and trail connections, are labeled and selected for analysis.

### 2.4 Location of Optimal Connections

Combinatorial optimization techniques were employed to identify and evaluate new street connections (see Fig. 2.), expanding on sidewalk siting analysis (Randall and Baetz, 2001). Fig. 2. A provides an example of residential parcels (orange nodes) within the Euclidean SWZ distance (blue dashed line) of the school (blue square) but not within the network SWZ distance. The residential parcels within the network SWZ distance are represented by green nodes and the residences outside of the Euclidean SWZ distance are the red nodes. In Fig. 2. B the dashed line depicts the optimal new connection that maximizes the number of residences now within the network SWZ distance and minimizes the length of the connection. To find the optimal thoroughfare that maximizes the number of residences within the network SWZ, exhaustive search optimization algorithms were performed in MATLAB (version 9.3 R2017b, see Algorithm 1). The exhaustive search optimization routine creates an edge for combinations of nodes and if that shortest path distance is less than the SWZ distance then it evaluates the cost of the new connection, i.e. the length of the new connection, and the benefit of the new connection, i.e. the number of residences that are now within the network SWZ distance (see the Supplemental Materials for the algorithm pseudocode). Note we parallelize the optimization routine as finding all of the solutions for large networks is computationally expensive (thousands of nodes results in millions of possible connections to evaluate). Heuristics have been developed to find near-optimal solutions when the exhaustive search is not computationally feasible (Auerbach,2018).

For the optimization algorithm, the network distance between two nodes is given by $d(i, j)$, and if the distance between a node and the school node ($S$) is less than the SWZ distance, $d(i, S) < D$ (where $D$ is 1 mile for elementary schools and 1.5 miles for middle schools) then the node $i$ is assigned to the set of close nodes ($N_C$), otherwise it is assigned to the set of distant nodes ($N_D$). Connections are made between two nodes, one being from $N_C$ and no restrictions on the other node. After a new connection is established, the residential nodes that are now within the network SWZ distance are assigned to the set $N_C$. The cost of the new connection is the connection length, $C(i, j) = d(i,j)$, and the benefit of this new connection $B(i,j)$ is the number of additional residences now within the SWZ. The optimal solution is the solution with the greatest benefit, or number of new residences now within the distance to the school, which can be expressed as the bi-objective function

$$O^* = \max_{(i,j)}(B(i,j) - C(i,j)).$$

## 2.5 Costs and Benefits of New Connections

The cost of a new connection was based on the average national cost for local or state governments to build a sidewalk (5 ft wide), approximately $1 million per mile, excluding right-of-way, crossing water/wetlands, and major topography (Bushell et al., 2013). The national average busing cost for a student was estimated to be approximately $1,000 in 2017 (using the most recent data from 2012-13 and adjusted for inflation) and used for the economic benefit of a new connection (National Center for Education Statistics, 2016). To estimate the health benefit of potential new street and trail connections, the number of additional residences within the SWZ were converted to the number of potential students who could walk or bike to school. The total benefit was calculated by taking this number of students and multiplying it by the estimated duration of the physical activity. The average U.S. student who walks to school acquires 16 minutes of moderate-intensity activity and they walk at a rate of 71.7 meters per minute (Crouter et al., 2013). Assuming a linear relationship between distance walked and walking time, it would take an average student 22.5 minutes to walk 1 mile, and this rate was used to calculate the walking benefit for the students along the optimal connection for each school evaluated.

# 3. Results

The location of the optimal connections are given in Fig. 4. The black square represents the school, and grey lines indicate streets. Green nodes are residences within the network SWZ distance, black nodes signify residences not within the network SWZ distance and orange nodes are the residences that are within the Euclidean SWZ distance after the new connection is made, denoted by the orange line. The length of the optimal connection, the number of residences and students included in the network SWZ with the new connection, and the total time spent walking to and from the school for these students are provided in Table 1. For each school the optimal student walking time was calculated by randomly selecting a number residences by the proportion of students per residence for the given school. A thousand random samples were used to calculate the average and standard deviation for the optimal walking time for each school.

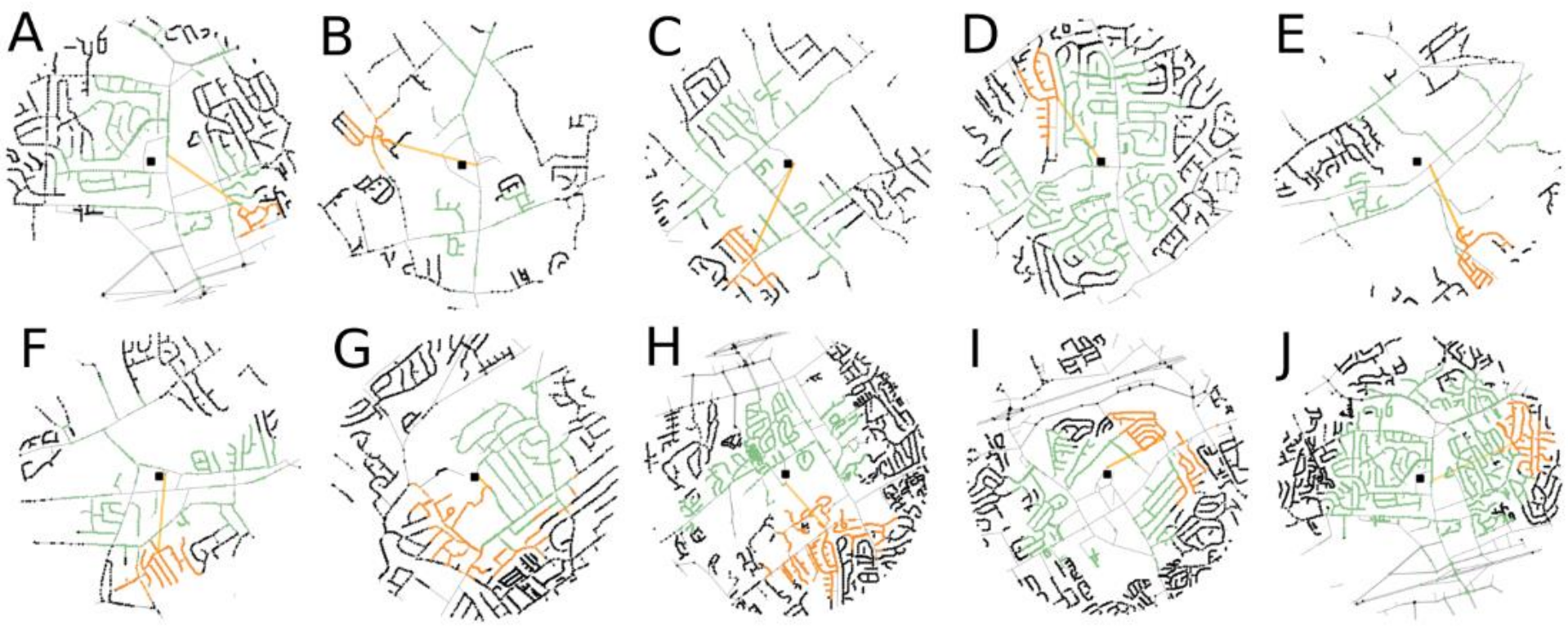

Fig. 4. Optimal connection results for the ten schools. The black square represents the school, and grey lines indicate streets. Green nodes are residences within the network SWZ distance, black nodes signify residences not within the network SWZ distance and orange nodes are the residences that are within the Euclidean SWZ distance after the new connection is made, denoted by the orange line.

Table 1. Active commuting characteristics and optimal results for the ten schools with the greatest walking potential. Schools denoted with an * are located in a suburban area, otherwise the school is located in a rural area. Standard deviations for the optimal walking times are given in the parentheses.

| | **Proportion of students** | | | | | | | | |
|---|---|---|---|---|---|---|---|---|---|
| **School** | **within Network SWZ** | **Euclidean SWZ** | **walking potential** | **# distant residences** | **students/ residence** | **optimal # residences** | **Optimal length, ft** | **Optimal # students** | **Optimal walking time, m (S.D.)** |
| Elementary | | | | | | | | | |
| A* | 0.11 | 0.28 | 292 | 1680 | 0.17 | 453 | 3704 | 77 | 2936 (30) |
| B | 0.01 | 0.03 | 239 | 1035 | 0.23 | 137 | 3891 | 39 | 1308 (16) |
| C | 0.05 | 0.18 | 228 | 1173 | 0.19 | 203 | 3613 | 39 | 1536 (24) |
| D* | 0.41 | 0.53 | 220 | 1911 | 0.12 | 255 | 3133 | 31 | 1168 (26) |
| E | 0.06 | 0.19 | 210 | 1154 | 0.18 | 287 | 3307 | 52 | 2008 (30) |
| F | 0.08 | 0.35 | 173 | 1120 | 0.15 | 287 | 2907 | 43 | 1502 (32) |
| G* | 0.20 | 0.31 | 171 | 1027 | 0.17 | 703 | 602 | 120 | 4220 (76) |
| Middle | | | | | | | | | |
| H* | 0.10 | 0.29 | 316 | 2033 | 0.13 | 940 | 3260 | 122 | 6452 (112) |
| I | 0.04 | 0.17 | 269 | 1058 | 0.25 | 464 | 3554 | 116 | 5994 (100) |
| J* | 0.29 | 0.45 | 209 | 4246 | 0.05 | 381 | 5027 | 19 | 1104 (28) |

## 4. Discussion

According to the results, increased connectivity could save the school system up to $120K/year in reduced busing costs for a one-time sidewalk cost as little as $114K, per school. These additional connections would also provide an elementary school students an additional 40 minutes, on average, of moderate-intensity physical activity (MVPA) from walking each school day (approximately 50 MVPA minutes gained for a middle school student) (Bassett et al., 2013). This is a significant amount of physical activity that could meet two-thirds of the national recommendations for physical activity (i.e., 60 minutes or more of MVPA per day) and help foster activity outside of commuting, reduce delinquencies, increase academic achievement, and reduce student BMI (U.S. Department of Health and Human Services, 2018).

Other additional benefits to increased active commuting to school, include reduced traffic, and its associated reduction in air pollution and accidents, and reduced healthcare costs. According to 2016 data, in the U.S. 1 in 5 school-age children are obese (Hales et al., 2017) and obese children ages 2-19 in the United States were found to stay on average 0.85 days longer for hospital treatment, incur $2000 additional charges, and $925 more in hospital costs than non-obese children (in 2017 dollars) (Trasande et al., 2009). Furthermore, obese children incurred $210 in higher outpatient visit expenditures, $124 higher prescription drug expenditures, and $13 higher emergency room expenditures (Trasande and Chatterjee, 2012). There were approximately 73.8 million children under the age of 18 in the United States in 2017 (Flood et al., 2018) and assuming obese children were hospitalized at the same rate as non-obese children (children were hospitalized in the U.S. at a rate of 0.0014 in 2012 (Witt et al., 2014)), this translates to an estimated $40 million dollars in additional hospital costs and $20 million in additional charges for the hospital treatments of obese children in 2017.

There are barriers to including walkabaility in residential development as Knox County, TN, like many other places in the U.S., has one group that make decisions about school siting and busing policies (the local school board) and another that makes decisions about the design of new neighborhoods and other developments (the local planning commission). Those groups may not always be aware of how decisions they make affect each other. These results describe and quantify the ways in which decisions made by the planning commission, specifically about thoroughfare connectivity in developments near schools, can impose additional costs on school systems, in the form of the need to provide more busing. Reduced thoroughfare connectivity around schools also imposes costs on families and on children themselves, in the form of less opportunity for physical activity, potentially lower academic achievement, and increased risk of obesity and other health problems.

It would be in the best interest of school systems to request that the planning commissions take into account the impact that low thoroughfare connectivity around schools has on busing costs. It is probably too late to remedy the low thoroughfare connectivity around the schools examined in this paper, because of cost and likely neighborhood opposition. But the key finding of this paper -- that short connections can vastly increase connectivity and walkability around schools, while decreasing future busing and health costs -- should inform future decision-making about neighborhood design around schools with the goal of reducing costs and improving students' health and academic achievement. An expanded working relationship between planning commissions and school systems may eventually lead to additional opportunities to collaborate toward cost savings, such as better coordination of school siting with other land use goals.

The methods provided here are an accessible approach to evaluate walkability for regional planners and local school officials. Results from this work have been shared with Knox County planning commissioners, the Knox County School Board, and other decision-makers with the intent of making them aware of the costs imposed by lack of thoroughfare connectivity in neighborhoods (Thaler and Sunstein, 2008). They could also influence school officials to consider thoroughfare connectivity and student active commuting when siting schools. Furthermore, these results could also start the dialogue of allowing greenways to be included as student commuting paths for school systems that currently do not consider them as permitted thoroughfares for student active commuting. The presence of greenways near residences has been shown to increase property values (Nichols and Crompton, 2005) as prospective home buyers are willing to pay more for a home in a walkable neighborhood (Knoxville Area Association of Realtors, 2017). After optimal connections are computationally identified, residents should be involved with the design of their future communities (Forester, 1999). Finally, school accessibility should be included in any research using an ecological systems approach or considering the built environment as a factor in childhood obesity (Fig. 5. modified from Davison and Birch, 2001).

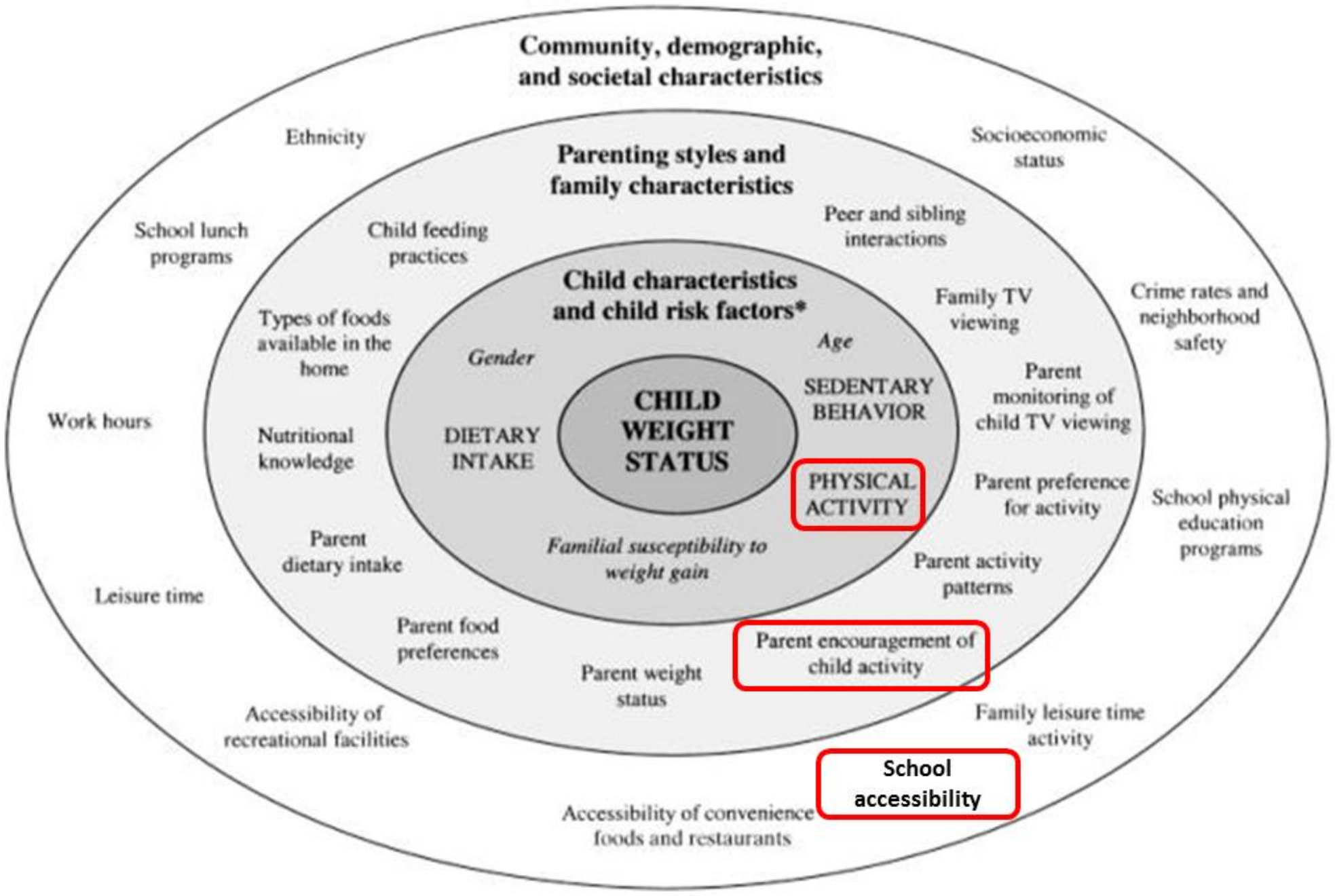

Fig. 5. Ecological systems theory approach to childhood obesity modified to include the accessibility of school (modified from Davison and Birch, 2001).

## 5. Limitations

There are several issues with the current analysis. The study area is limited to one U.S. county, and to make these results more robust, the networks for schools in additional study areas should be evaluated. There are several methodological issues. Residences may be placed on the nearest street but that may not be the street they are physically located on. The error rate for this occurring is small for this study area but could vary for different study areas. Restrictions on the locations of new streets and trails was not included. For example, thoroughfares cannot be built on slopes beyond a specific threshold, and it may be cost prohibitive or impossible to route them through specific parcel types (such as commercial areas, mining sites and landfills) or highways. Solutions are dependent on the location of nodes at every parcel and intersection. This can lead to large sparse areas of possible connectivity and could be corrected with the placement of artificial nodes at regular intervals at the expense of computational costs. The issue that newly constructed connections can intersect existing streets (planarity) was not resolved in this study. Calculating the determinants for each pair of lines is computationally expensive. Lastly and due to the limitations of shapefiles, it is recommended that government agencies move away from their dependence on them for spatial data. There are several emerging data formats, such as OGC GeoPackage and GeoJSON, that are robust and free.

## 6. Conclusions

The results show that even the addition of short thoroughfares can significantly increase the number of residences within the walking zone of a school. These additional residences now within the walking zone increase the potential number of students who could actively commute to school, and commuting is positively correlated with improved health and education achievement. Planners and other decision-makers can use this result to increase their awareness about the costs, to families and to school systems, imposed by the lack of thoroughfare connectivity in new neighborhoods. We offer an approach to evaluate school walkability for regional planners, school systems, and health departments.

## Acknowledgments

We would like to thank Alex Zendel (GIS Analyst at the Knoxville-Knox County Metropolitan Planning Commission) for compiling the school network data, and Rickey Grubb (Director of Enrollment and Transportation at Knox County Schools) for relevant information about school busing costs.

**Contributions: Jeremy Auerbach:** Conceptualization, Methodology, Software, Validation, Formal Analysis, Investigation, Data Curations, Writing - Original Draft, Writing – Review & Editing, Visualization **Eugene Fitzhugh:** Conceptualization, Writing - Original Draft, Writing – Review & Editing, Funding Acquisition **Ellen Zavisca:** Conceptualization, Writing - Original Draft, Writing – Review & Editing

**Funding:** JDA and ECF were supported by a University of Tennessee Transportation Center Fellowship award.

## References

J. Auerbach. Essays in network theory applications for transportation planning. PhD thesis, University of Tennessee, 2018.

S. Babey, T. Hastert, W. Huang, and E. Brown. Sociodemographic, family, and environmental factors associated with active commuting to school among US adolescents. Journal of Public Health Policy, 30(Suppl 1):S203-S220, 2009. doi: 10.1057/jphp.2008.61.

D. R. Bassett, E. C. Fitzhugh, G. W. Heath, P. Erwin, F. G. M., D. L. Wolf, W. A. Welch, and A. B. Stout. Estimated energy expenditures for school-based policies and active living. American Journal of Preventive Medicine, 44(2):108-113, 2013. doi: 10.1016/j.amepre.2012.10.017.

T. Bungum, M. Lounsbery, S. Moonie, and J. Gast. Prevalence and correlates of walking and biking to school among adolescents. Journal of Community Health, 34(2):129-134, 2009. doi: 10.1007/s10900-008-9135-3.

M. Bushell, B. Poole, D. Rodriguez, and C. Zegeer. Costs for pedestrian and bicyclist infrastructure improvements: A resource for researchers, engineers, planners and the general public. Technical report, UNC Highway Safety Research Center, 2013.

Centers for Disease Control and Prevention. Barriers to children walking to or from school. United States, 2004. Technical report, U.S. Department of Health and Human Services, 2005.

Centers for Disease Control and Prevention. The association between school based physical activity, including physical education, and academic performance. Technical report, U.S. Department of Health and Human Services, 2010.

J. F. Chriqui, D. R. Taber, S. J. Slater, T. L., K. M. Lowrey, and F. J. Chaloupk. The impact of state safe routes to school-related laws on active travel to school policies and practices in U.S. elementary schools. Health & Place, 18(1):8-15, 2012. doi: 10.1016/j.healthplace.2011.08.006.

C. Coughenour, S. Clark, A. Singh, and J. Huebner. Are single entry communities and cul-de-sacs a barrier to active transport to school in 11 elementary schools in Las Vegas, NV metropolitan area? Preventive Medicine Reports, 6:144-148, 2017. doi: 10.1016/j.pmedr.2017.02.011.

S. E. Crouter, M. Horton, and D. R. Bassett. Validity of ActiGraph child-specific equations during various physical activities. Medicine & Science in Sports & Exercise, 45(7):1403-1409, 2013. doi: 10.1249/MSS.0b013e318285f03b.

K. Davison and L. Birch. Childhood overweight: a contextual model and recommendations for future research. Obesity Reviews, 2(3):159-171, 2001.

A. DeNisco. School bus driver shortage drives new incentives, October 2015. https://www.districtadministration.com/article/school-bus-driver-shortage-drives-new-incentives.

S. Flood, M. King, R. Rodgers, S. Ruggles, and J. R. Warren. Integrated Public Use Microdata Series, Current Population Survey: Version 6.0 [dataset], 2018.

J. F. Forester. The Deliberative Practitioner: Encouraging Participatory Planning Processes. MIT Press, 1999.

L. D. Frank, J. F. Sallis, T. L. Conway, J. E. Chapman, B. E. Saelens, and W. Bachman. Many pathways from land use to health: Associations between neighborhood walkability and active transportation, body mass index, and air quality. Journal of the American Planning Association, 72(1):75-87, 2006. doi: 10.1080/01944360608976725.

B. Giles-Corti, G. Wood, T. Pikora, V. Learnihan, M. Bulsara, K. Van Niel, A. Timperio, G. McCormack, and K. Villanueva. School site and the potential to walk to school: the impact of thoroughfare connectivity and traffic exposure in school neighborhoods. Health & Place, 17(2):545-550, 2011. doi: 10.1016/j.healthplace.2010.12.011.

I. Grafova. Overweight children: assessing the contribution of the built environment. Preventive Medicine, 47(3):304-308, 2008. doi: 10.1016/j.ypmed.2008.04.012.

C. M. Hales, M. D. Carroll, C. D. Fryar, and C. L. Ogden. Prevalence of obesity among adults and youth: United States, 2015-2016. Technical report, National Center for Health Statistics, 2017.

H. J. Kim and C. Lee. Does a more centrally located school promote walking to school? Spatial centrality in school-neighborhood settings. Journal of Physical Activity and Health, 13:481-487, 2016. doi: 10.1123/jpah.2015-0221.

Knox County Schools Transportation Department. Bus rider eligibility: Parent Responsibility Zone (PRZ). Technical report, Knox County Schools, 2009.

Knoxville Area Association of Realtors. Walkability survey. Technical report, Knoxville Area Association of Realtors, 2017.

C. Kouri. Wait for the bus: How low country school site selection and design deter walking to school and contribute to urban sprawl. Technical report, South Carolina Coastal Conservation League, 1999.

K. Larsen, J. Gilliland, P. Hess, P. Tucker, J. Irwin, and M. He. The influence of the physical environment and sociodemographic characteristics on children's mode of travel to and from school. American Journal of Public Health, 99(3):520-526, 2009. doi: 10.2105/AJPH.2008.135319.

D. R. Lubans, C. A. Boreham, P. Kelly, and C. E. Foster. The relationship between active travel to school and health-related fitness in children and adolescents: a systematic review. International Journal of Behavioral Nutrition and Physical Activity, 8(5), 2011. doi: 10.1186/1479-5868-8-5.

W. E. Marshall, D. P. Piatkowski, and N. W. Garrick. Community design, street networks, and public health. Journal of Transport & Health, 1(4):326-340, 2014. doi: 10.1016/j.jth.2014.06.002.

N. C. McDonald. Active transportation to school: Trends among U.S. schoolchildren, 1969-2001. American Journal of Preventive Medicine, 32:509-516, 2007. doi: 10.1016/j.amepre.2007.02.022.

N. C. McDonald and A. E. Aalborg. Why parents drive children to school: Implications for Safe Routes to School programs. Journal of the American Planning Association, 75(3): 331-342, 2009. doi: 10.1080/01944360902988794.

N. C. McDonald, A. L. Brown, L. M. Marchetti, and M. S. Pedroso. U.S. school travel, 2009 an assessment of trends. American Journal of Preventive Medicine, 41:146-151, 2011. doi: 10.1016/j.amepre.2011.04.006.

J. A. Mendoza, K. Watson, N. Nguyen, E. Cerin, T. Baranowski, and T. A. Nicklas. Active commuting to school and association with physical activity and adiposity among US youth. Journal of Physical Activity & Health, 8(4):488-495, 2011.

J. A. Mendoza, D. Cowan, and Y. Liu. Predictors of children's active commuting to school: an observational evaluation in 5 U.S. communities. Journal of Physical Activity & Health, 11(4):729-733, 2014. doi: 10.1123/jpah.2012-0322.

P. R. Nader, R. H. Bradley, and R. M. Houts. Moderate-to-vigorous physical activity from ages 9 to 15 years. Journal of the American Medical Association, 300:295-305, 2008. doi: 10.1001/jama.300.3.295.

National Center for Education Statistics. Digest of Education Statistics. Technical report, Institute of Education Sciences, 2016.

National Center for Statistics and Analysis. 2015 motor vehicle crashes. Technical report, National Highway Traffic Safety Administration, 2016.

N. M. Nelson, E. Foley, D. J. O'Gorman, N. M. Moyna, and C. B. Woods. Active commuting to school: How far is too far? International Journal of Behavioral Nutrition and Physical Activity, 5(1), 2008. doi: 10.1186/1479-5868-5-1.

S. Nichols and J. L. Crompton. The impact of greenways on property values: Evidence from Austin, Texas. Journal of Leisure Research, 37(3):321-341, 2005. doi: 10.1080/00222216.2005.11950056.

Office of Children's Health Protection. School siting guidelines. Technical report, United States Environmental Protection Agency, 2011.

N. Oreskovic, K. Kuhlthau, D. Romm, and J. Perrin. Built environment and weight disparities among children in high- and low-income towns. Academic Pediatrics, 9(5):315-321, 2009. doi: 10.1016/j.acap.2009.02.009.

Pedestrian and Bicycle Information Center. Safe routes to school guide. Technical report, University of North Carolina Highway Safety Research Center, 2010.

C. A. Perry. The neighborhood unit, a scheme of arrangement for the family-life community. In Monograph 1 of Regional Plan of N.Y. Regional survey of N.Y. and its environs, volume 7, pages 2-140. 1929. Reprinted Routledge/Thoemmes, London, 1998, 25-44.

A. E. Price, D. M. Pluto, O. Ogoussan, and J. A. Banda. School administrators' perceptions of factors that influence children's active travel to school. Journal of School Health, 81 (12):741-748, 2011. doi: 10.1111/j.1746-1561.2011.00653.x.

T. A. Randall and B. W. Baetz. Evaluating pedestrian connectivity for suburban sustainability. Journal of Urban Planning and Development, 127(1):1-15, 2001. doi: 10.1061/0733-9488.

Robert Wood Johnson Foundation. Active education: Physical education, physical activity and academic performance. Technical report, Robert Wood Johnson Foundation, 2009.

J. R. Sirard and M. E. Slater. Walking and bicycling to school: A review. American Journal of Lifestyle Medicine, 32:372-396, 2008. doi: 10.1177/1559827608320127.

J. Spence, N. Cutumisu, J. Edwards, and J. Evans. Influence of neighbourhood design and access to facilities on overweight among preschool children. International Journal of Pediatric Obesity, 3(2):109-116, 2008. doi: 10.1080/17477160701875007.

S. Tewahadea, K. Lia, R. B. Goldstein, D. Haynie, R. J. Iannotti, and B. Simons-Morton. Association between the built environment and active transportation among U.S. adolescents, 15, 2019. doi: 10.1016/j.jth.2019.100629.

R. H. Thaler and C. R. Sunstein. Nudge: Improving Decisions about Health, Wealth, and Happiness. Yale University Press, 2008.

The National Center for Safe Routes to School. Trends in walking and bicycling to school from 2007 to 2014. Technical report, U.S. Department of Transportation Federal Highway Administration, 2016.

Transportation Consultants. Walk-to-school prioritization analysis for the schools of Knox County, Tennessee. Technical report, Knox County Department of Engineering & Public Works, 2014.

L. Trasande and S. Chatterjee. The impact of obesity on health service utilization and costs in childhood. Obesity, 17(9):1749-1754, 2012. doi: 10.1038/oby.2009.67.

L. Trasande, Y. Liu, G. Fryer, and M. Weitzman. Effects of childhood obesity on hospital care and costs, 1999-2005. Health Affairs, 28(4):751-760, 2009. doi: 10.1377/hlthaff.28.4.w751.

G. Turrell, B. Hewitt, J. Rachele, B. Giles-Corti, L. Busija, and W. Brown. Do active modes of transport cause lower body mass index? Findings from the HABITAT longitudinal study. Journal of Epidemiology and Community Health, 72:294-301, 2018. doi: 10.1136/jech-2017-209957.

U.S. Department of Health and Human Services. Physical activity guidelines for Americans, 2nd edition. Technical report, U.S. Department of Health and Human Services, 2018.

G. Welk. Cardiovascular fitness and body mass index are associated with academic achievement in schools. Technical report, Cooper Institute, 2009.

W. P. Witt, A. J. Weiss, and A. Elixhauser. Overview of hospital stays for children in the United States, 2012. Technical report, Agency for Healthcare Research and Quality, 2014.

## Supplemental Materials

## Algorithm Psuedocode

Exhaustive search pseudocode

| | |
|---|---|
| **for** $i$ **in** $N_D$ **do** | Select a distant node |
|   **for** $j$ **in** $N_C$ **do** | Select a close node |
|     **if** $d(i,j) + d(j,S) < D$ **then** | If the new connection changes the distant node to close |
|       $C(i,j) = d(i,j)$ | Calculate the distance of the new connection |
|       **for** $k$ **in** $N_D$ **do** | Select a distant residence |
|         **if** $d(k,i) + d(i,j) + d(j,S) < D$ **then** | Calculate the distant to the school |
|           $k \epsilon N_C'$ | If the distant residence is within the SWZ, then label it close |
|         **end if** | |
|       **end for** | |
|       $B(i,j) = \lvert N_C' \rvert$ | Calculate the number of new close residences |
|     **end if** | |
|   **end for** | |
| **end for** | |

## Data Engineering

The spatial data provided by KGIS was in shapefiles, which are commonly used by government agencies but have significant limitations for analysis. The data cleaning and processing methods described below are provided for planners and researchers to use for their own systems (see Karduni et al., 2016, and Oliver et al., 2007, for additional useful methods to process shapefile data for network analysis).

To identify optimal thoroughfares a network approach was used and the data was converted to a network with nodes at residences and street intersections and the streets were converted to edges. To accomplish this data transformation, in ArcMap (version 10.6):

1. buffers were created around each school (1 mile for elementary schools and 1.5 miles for middle schools).
2. street edge and street intersection node development:
    a. street data outside of the school buffer were removed (using the clip tool),
    b. highways were removed, as students will not use them for active commuting,
    c. streets part of the school property were also removed (as schools do not want direct connections),
    d. nodes at the street intersections were created and labeled as intersections,
        i. points at the street intersections were created (intersect tool with output type point),
        ii. two numeric fields were added to this layer using a field type DOUBLE to ensure maximum precision and the spatial coordinates were identified for the street intersection points (calculate geometry),
        iii. ArcGIS creates these intersections as multipoint geometries (coinciding with the number of streets that meet at the intersection), therefore these intersections were converted to single points (feature to point tool) the

identical intersection points were removed by their coordinates (delete identical tool),

iv. a numeric field was added to this layer and all street intersection points were given a value of 0,

3. residential node development1[1]:
    i. residential parcels outside of the school buffer were removed (using the clip tool),
    ii. vacant residential lots were removed (select by attributes),
    iii. nodes at the street intersections were created and labeled as intersections,
        i. nearest street location was identified for each residence (the near tool),
        ii. the attributes of these points were converted to a database table (output attribute as .dbf),
        iii. these residential coordinates plotted and saved as a layer (display XY data with the Near_X and Near_Y coordinates),
        iv. a numeric field was added to the layer and all residential points were given a value of 1,
4. a school node was placed at the nearest street intersection and labeled,
5. the street intersection point data, the school point, and the residential point data were merged (merge tool) with the following important fields kept: (1) X-coordinate, (ii) Y-coordinate, and (iii) node type and exported as a text file,
6. the street polylines were cut at each residential and intersection point (split line at point tool). A search radius of 100 ft was used to offset any residential nodes that were not exactly on a street and this also provides additional solutions for the optimization method.

Due to the limitations of ArcGIS, for example the inability of the software to produce a network connectivity matrix (which is an extremely useful mathematical representation), the cut street data files were imported into Python and the NetworkX package (version 2.1) was used to produce adjacency (connectivity) matrices and a vector of connection lengths (Hagberg, 2017). These adjacency matrices, edge lengths, and the node data (schools, residences, and intersections) were imported into MATLAB (version 9.3 R2017b) for the optimization routines (Karduni et. al, 2016). Network distances between all nodes were calculated and nodes were labeled as close or distant based on the walking network distance.

Karduni A, Kermanshah A, Derrible S. A Protocol to Convert Spatial Polyline Data to Network Formats and Applications to World Urban Road Networks. Scientific Data. 2016;3. https://doi.org/10.1038/sdata.2016.46

Oliver LN, Schuurman N, Hall AW. Comparing Circular and Network Buffers to Examine the Influence of Land use on Walking for Leisure and Errands. International Journal of Health Geographics. 2007;6(41). https://doi.org/10.1186/1476-072X-6-41.

Hagberg A. NetworkX. https://networkx.github.io/. Published 2017.

[1] Apartments were already divided into separate single family residence points. If this is not the case then apartments should be labeled with the number of units for the optimization method to consider when calculating the benefit of the new connections.